# Unusual Magnetic Properties in Layered Magnetic Topological Insulator EuSn$_2$As$_2$

Huijie Li,[1,2] Wenshuai Gao,[1,†] Zheng Chen,[2,3] Weiwei Chu,[2,3] Yong Nie,[2,3] Shuaiqi Ma,[1] Yuyan Han,[2] Min Wu,[2] Tian Li,[2] Qun Niu,[2] Wei Ning,[2] Xiangde Zhu,[2,†] and Mingliang Tian[1,2,4]

[1]*Institutes of Physical Science and Information Technology, Anhui University, Hefei 230601, China*

[2]*Anhui Key Laboratory of Condensed Matter Physics at Extreme Conditions, High Magnetic Field Laboratory, HFIPS, Anhui, Chinese Academy of Sciences, Hefei 230031, P. R. China*

[3]*Department of Physics, University of Science and Technology of China, Hefei 230031, Anhui, China*

[4]*School of Physics and Materials Science, Anhui University, Hefei 230601, China*

## Abstract

EuSn$_2$As$_2$ with layered rhombohedral crystal structure is proposed to be a candidate of intrinsic antiferromagnetic (AFM) topological insulator. Here, we have investigated systematic magnetoresistance (MR) and magnetization measurements on the high quality EuSn$_2$As$_2$ single crystal with the magnetic field both parallel and perpendicular to (*00l*) plane. Both the kink of magnetic susceptibility and longitudinal resistivity reveal that EuSn$_2$An$_2$ undergoes an AFM transition at $T_N$ = 21 K. At $T$ = 2 K, the magnetization exhibits two successive plateaus of ~ 5.6 $\mu_B$/Eu and ~ 6.6 $\mu_B$/Eu at the corresponding critical magnetic fields. Combined with the negative longitudinal MR and abnormal Hall resistance, we demonstrate that EuSn$_2$An$_2$ undergoes complicated magnetic transitions from an AFM state to a canted ferromagnetic (FM) state at $H_c$ and then to a polarized FM state at $H_s$ as the magnetic field increase.

† Corresponding author email: gwsh@ahu.edu.cn; xdzhu@hmfl.ac.cn;

## I. INTRODUCTION

The quantum anomalous Hall (QAH) effect was first realized in the magnetically doped thin (Bi, Sb)$_2$Te$_3$ topological insulator (TI) films at extremely low temperatures due to the inhomogeneous magnetic impurities [1]. A lot of theoretical and experimental efforts have been made to explore the stoichiometric TIs with an inherent magnetic order, i.e., intrinsic magnetic topological insulator [2-6]. Recently, extensive works were focused on the layered tetradymite compound MnBi$_2$Te$_4$ and its related materials. MnBi$_2$Te$_4$ is an intrinsic A-type antiferromagnetic (AFM) TI with Mn ferromagnetic (FM) layers stacked antiferromagnetically along the *c*-axis [7]. A great deal of theoretical and experimental works reveal rich quantum states in MnBi$_2$Te$_4$ thin flakes with different thickness, including QAH effect, axion insulating state, high-Chern-number and high-temperature Chern insulation state [8-11]. These findings demonstrate that exploring intrinsic magnetic TIs is an effective way to study and realize novel magnetic topological effects.

Except for layered tetradymite compound MnBi$_2$Te$_4$, the intrinsic magnetic topological insulators are still rare. The Eu-based magnetic materials, including EuSn$_2$As$_2$, EuIn$_2$As$_2$ and EuSn$_2$P$_2$, are one family of the intrinsic magnetic topological insulators that have been theoretically predicted and then confirmed by spectroscopy studies [7, 12-14]. Their magnetic ground states at low temperatures are mostly A-type AFM state like MnBi$_2$Te$_4$, i.e., the magnetic moments are ferromagnetically aligned in the layer but antiferromagnetically coupled interlayer [7, 13-15]. The interplay between ordered magnetic state and the topological electronic state are expected to induce abundant novel quantum states. For example, EuIn$_2$As$_2$ is theoretically proposed to be an axion insulator with AFM long-range order, and its topological character has been revealed experimentally [12, 16, 17]. An interesting Weyl semimetal state induced by magnetic field is also demonstrated in the paramagnetic (PM) phase of EuCd$_2$As$_2$ [18]. Among these compounds, EuSn$_2$As$_2$ with a space group of *R-3m* displays a unique advantage in realizing quantum effects or devices [7, 19]. Previous studies have revealed that EuSn$_2$As$_2$ undergoes a PM to AFM phase transition at about $T_N$ = 21 K, then the system enters into AFM TI state with an obvious Dirac surface state. While the previous theoretical and experimental works have promoted the research progress of Eu-based magnetic topological insulators, many novel quantum states, such as the quantum Hall effect and the axion insulating state which have been identified in MnBi$_2$Te$_4$ family, have not been experimentally observed. One possible reason is that the magnetic moments of Eu 4$f$ states, where the magnetism of Eu-based magnetic topological insulators results from, localize far below Fermi level $E_F$ and coupled weakly with the nontrivial topologic electron states [15]. However, there are also theories indicate that when the local spin of the Eu$^{2+}$ ions enter the Zintl phase of the intermetallic compound, the Dirac surface state of the three-dimensional topological insulator can work as the medium of the Ruderman-Kittel-Kasuya-Yosida (RKKY) long-range ferromagnetic exchange effect, which enhances the magnetic coupling effect and is conducive to the realization of topological quantum phenomena [19].

Herein, we performed systematic magnetotransport measurements on high quality EuSn$_2$As$_2$ single crystals with different magnetic field orientations. The magnetic susceptibility verifies the AFM ground state with transition temperature $T_N$ = 21 K. At 2 K, the magnetization curves show the first plateau of ~ 5.6 $\mu_B$/Eu atom and the second magnetization plateau of ~ 6.6 $\mu_B$/Eu atom with $H$ up to 14 T. Furthermore, the unusual magnetization plateaus are accompanied by abnormal MR at the critical magnetic fields. All the results help us to understand the successive magnetic transitions from an AFM state to a canted FM state followed by a polarized FM state for higher magnetic fields.

## II. EXPERIMENTAL PREPARATION

EuSn$_2$As$_2$ single crystals were grown via the self-flux method, which is different from the solidification of the stoichiometric melt process reported by Arguilla *et al.* [19]. Eu shots, Sn lumps, and As pieces in a molar ratio of 1:10:2 as the starting materials were put in an alumina crucible and sealed in an evacuated quartz tube. The tube was heated up to 750 °C in 33 h, held for 48 h, and then slowly cooled to 400 °C at a rate of 2 ºC/h. After a centrifugal separation operation at this temperature, argenteous single crystals with regular dimension of about 0.5×1.25×0.2 mm$^3$ were obtained, as shown in the inset of Fig. 1(b). The self-flux method introduces almost no impurity elements. The possible residual tiny Sn attached on the single crystal surface can be removed by mechanical exfoliation, and has no effect on the experimental results. The compositions and elemental stoichiometry of crystals were determined by energy-dispersive x-ray spectrometer (EDS) on Oxford X-MAX electron spectrometer equipped on the focused electron beam-ion beam dual beam system (Helios Nanolab600i, FEI Inc.).

X-ray diffraction (XRD) pattern measurements were carried out upon a Rigaku-TTR3 X-ray diffractometer using the high intensity graphite monochromatized Cu Kα radiation. Magnetization measurements were carried out upon a Magnetic Property Measurement System (MPMS3-7 T, Quantum Design Inc.) and a Physical Property Measurement System (PPMS-16T, Quantum Design Inc.) equipped with a vibrating sample magnetometer (VSM) option. Electrical transport measurements were carried out upon Physical Property Measurement System DynaCool (PPMS-14 T, Quantum Design Inc.) from 2 K to 300 K. The six-probe electrodes with a standard Hall bar configuration were prepared using electron-beam lithography (EBL) at focused ion beam dual beam system (Helios Nanolab600i, FEI Inc.) equipped with the Rathe Multibeam Nanopattern Generator, followed by Ti (5nm)/Au (100 nm) deposition as the contacts. The scanning electron microscope (SEM) image of the device is shown in the inset of Fig. 2b.

### III. RESULTS AND DISCUSSION

XRD data of EuSn$_2$As$_2$ powder were analyzed using the EXPGUI-GSAS Rietveld refinement method [20, 21]. The initial lattice parameters from Arguilla *et al* were adopted during the process of refinement. [19]. Fig.1(a) displays the refined result of the XRD pattern, where the experimental results are in good agreement with the calculated results. Through the Rietveld refinement, the lattice parameters are identified as $a = b = 4.197$ (2) Å, $c = 26.393$ (8) Å and $α = β = 90º$, $γ = 120º$, which are consistent with the previous report [19]. According to these parameters, we draw the EuSn$_2$As$_2$ crystal structure using CrytalMaker software, as shown in the inset of Fig. 1(a). $EuSn_2As_2$ presents a layered structure with space group *R-3m* (166), in which two SnAs layers intercalate with one Eu layer and form a two-dimensional (2D) sheet. The two adjacent SnAs layers are weakly coupled via the van der Waals force, which makes $EuSn_2As_2$ crystals to be easily exfoliated using silicone-free tape like $MnBi_2Te_4$ [7]. Fig. 1(b) shows the XRD pattern of $EuSn_2As_2$ single crystal, all the sharp diffraction peaks can be well indexed to the (*00l*) plane and reveal excellent crystallinity. The EDS measurements also establish the chemical composition ratio of Eu : Sn : As = 19.86 : 40.00 : 40.15 to be nearly stoichiometric ratio. All the results confirm the high quality of the $EuSn_2As_2$ crystals.

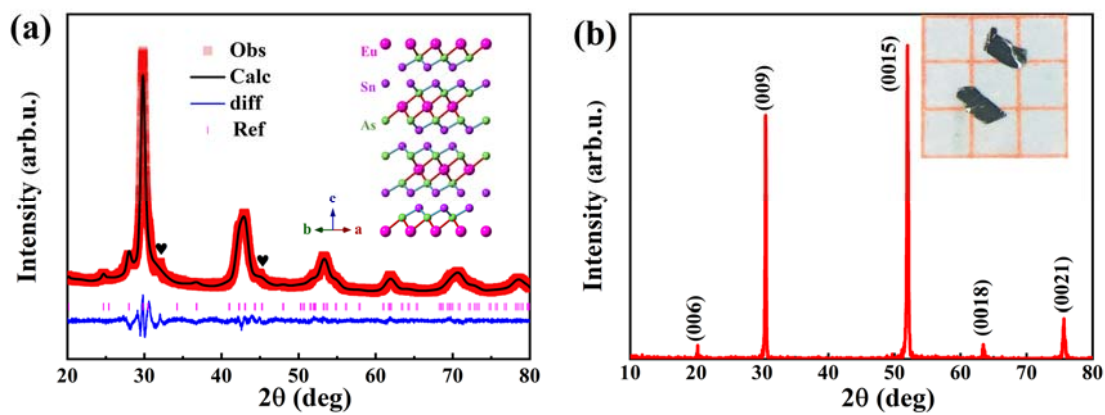

FIG. 1. (a) Rietveld refinement for $EuSn_2As_2$ powder. The weak peaks marked with "♥" are results from the surface flux Sn. The inset shows the structural illustration of $EuSn_2As_2$. (b) XRD pattern of $EuSn_2As_2$ single crystal with only (*00l*) peaks related to the rhombohedral $EuSn_2As_2$ phase. The inset is an optical photograph of as-grown $EuSn_2As_2$ single crystals.

To get insight into the magnetic properties of $EuSn_2As_2$, Fig. 2(a) shows the temperature-dependent magnetic susceptibility with an applied magnetic field of 0.05 T for *H//ab* plane and *H//c* axis, respectively. Zero-field-cooling (ZFC) and field-cooling (FC) cycles are nearly superimposable. The curve shows a kink at about 21 K for both magnetic field orientations, which indicates an AFM transition. Below 10 K, the $\chi$(T) curve presents an unusual rapidly increase for *H//ab* plane, which is consistent with the previous report [7]. This behavior may result from the local disorder of magnetic moments. The dotted lines are the inverse susceptibility, $\chi^{-1}$, as a function of temperatures. For $H//ab$ plane, the curve can be well fitted using the Curie-Weiss law at temperatures above 30 K. The best fitting yields the Weiss constant ($\Delta$) of 17.5 K and the effective magnetic moment of $\mu_{eff} = 7.74\ \mu_B$, which is slightly smaller than the theoretical value for the divalent state (7.94 $\mu_B/Eu^{2+}$). Fig. 2(b) shows the resistance versus temperature curves of EuSn$_2$As$_2$ bulk single crystal and 600 nm thick flake, respectively. Both samples show a metallic behavior in high temperature range and then display an anomalous peak at about 21 K. Such an anomalous resistance kink is attributed to the Kondo scattering, which has been proposed in the previous reports [15, 19].

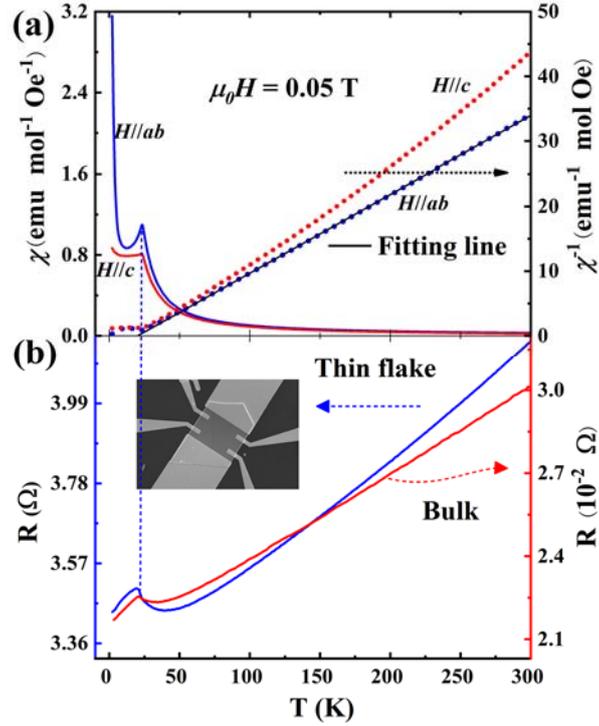

FIG. 2. (a) Temperature-dependent magnetic susceptibility, $\chi$, of EuSn$_2$As$_2$ for $H//ab$ plane and $H//c$ axis, respectively, measured at 0.05 T. The dotted lines show the $\chi^{-1}$ versus T curves, and the black solid line is the fitting result using the Curie-Weiss law. (b) The $R$-$T$ curves obtained from EuSn$_2$As$_2$ bulk crystal and nanoflake. The inset is the SEM image of the EuSn$_2$As$_2$ device with a six-probe configuration.

Fig. 3(a) presents the magnetic field dependence of magnetizations with magnetic field parallel to the *c*-axis at various temperatures in the field range from -14 T to 14 T. At $T$ = 2 K, the magnetization increases linearly with increasing magnetic field up to 4.6 T, and then presents an apparent plateau for *H//c*, where the critical field of 4.6 T is consistent with the previous reports [7, 19, 15]. However, such a magnetization plateau with ~ 5.6 $\mu_B$/Eu atom tends to rise with further increase the magnetic field. In our study, the applied external magnetic field *H* is up to 14 T. Interestingly, a second plateau with ~ 6.6 $\mu_B$/Eu atom appears around 10 T at low temperatures, then gradually smears out with the increase of temperature. The enlarged plateau region is shown in the inset of Fig.3(a), and the magnetization value at the second plateau is slightly smaller than the theoretical value of 7.0 $\mu_B$/Eu$^{2+}$ for the divalent states. Our data clearly demonstrate that EuSn$_2$As$_2$ undergoes successive magnetic transitions induced by magnetic field. Such successive magnetization plateaus have been reported in many rare-earth compounds including some Eu-based compounds [22-24].

Fig.3(b) shows the MR as a function of magnetic field with *H//c* at different temperatures. At $T$ = 80 K, it displays a positive MR in the field range of -14 T to 14 T. Below $T$ = 60 K, negative MRs are emerged and further enhanced as the temperature decreases. At $T$ = 2 K, a distinct negative MR as large as 5.7% is observed around the critical field of the first magnetization plateau $H_c$ = 4.6 T. Then the MR curve turns up with a positive slope as the field increases followed by a kink at the field, $H_s$ = 10 T, of the second magnetization plateau. The blue dotted lines in Fig.3 (a) and 3(b) clearly mark the corresponding critical magnetic fields.

To further shed light on the nature of the magnetism in EuSn$_2$As$_2$, Fig. 3(c) presents the field-dependent magnetization with *H//ab*-plane and *H//c* axis at 2 K, respectively. The first critical fields are identified to be about $H_c$ = 3.2 T and 4.6 T for *H//ab*-plane and *H//c* axis, respectively, indicating a considerable magnetic anisotropy in EuSn$_2$As$_2$. No obvious magnetization hysteresis is observed by cycling magnetic fields from 14 T to -14 T. This anisotropic property is also found by the MR measurement, as shown in Fig.3(d). According to previous reports and our susceptibility measurements, an A-type AFM order characterized by the in-plane FM layers stacked antiferromagnetically along the *c*-axis in EuSn$_2$As$_2$ can be qualitatively determined below $T_N$ = 21 K [7,15,19].

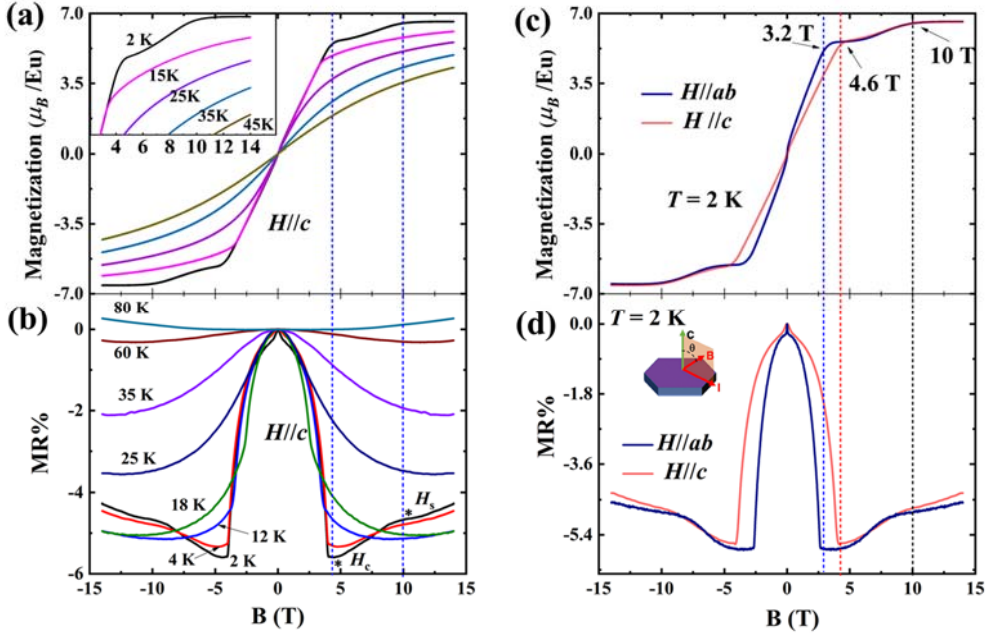

FIG. 3. (a) and (b) are the magnetization and the MR curves with *H//c* axis at various temperatures. The inset is a close-up of the *M-H* curves near $H_c$ and $H_s$. (c) and (d) are the magnetization and MR curves at 2 K with *H//c* axis and *H//ab*-plane, respectively.

Fig.4 shows the Hall resistivity versus magnetic fields at different temperatures. In comparison with the previous report [15], where the Hall resistivity is totally linearly dependent on the magnetic field over the entire temperature range, here we observed an unambiguous kink of Hall resistivity around $H_c$ = 3.2 T for *H//c* below $T_N$ = 21 K. The anomalous Hall effect in EuSn$_2$As$_2$ may result from the coupling between the localized magnetic ordering and the conduction electronic states, which provides positive signal for further exploration of novel quantum effects. By analyzing the Hall resistivity at high magnetic fields, the hole-type carrier concentration and mobility are estimated to be $n$ = 4.29 × 10$^{20}$ cm$^{-3}$ and $\mu$ = 70.03 cm$^2$V$^{-1}$s$^{-1}$, respectively. Such a carrier concentration is also comparable to a previous report [15]. The Hall resistivity at high field range is almost temperature independent, demonstrating that the AFM transition at 21 K has less dependent on the changes of electronic states.

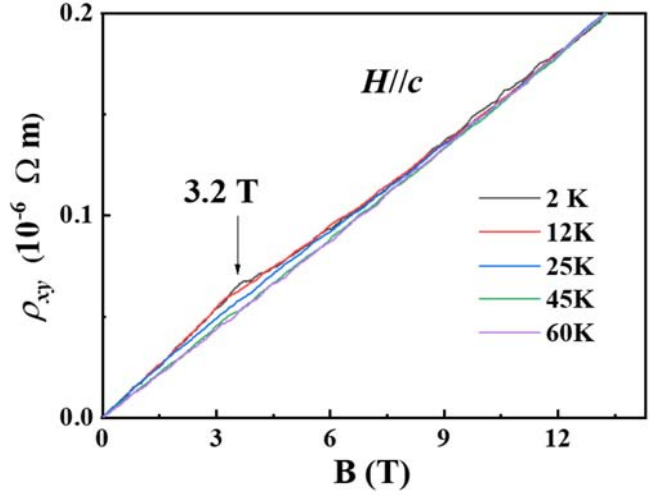

FIG. 4. The Hall resistivity $\rho_{xy}$ at various temperatures with applied field from 0 to 14 T.

The successive magnetization plateaus observed in EuSn$_2$As$_2$ provide us more opportunities to study on its complicated magnetic structures. Firstly, the magnetic momentum of ~ 5.6 $\mu_B$/Eu atom at the first critical filed $H_c$ is quite smaller than the theoretical value of 7 $\mu_B$/Eu$^{2+}$, which clearly reveals a magnetic field induced magnetic transition from AFM state to canted FM state between $H_c$ and $H_s$. Besides, even if the applied magnetic field is much larger than the second critical magnetic field $H_s$ =10 T, the magnetic momentum of ~ 6.6 $\mu_B$/Eu atom at the second magnetization plateau is very close but still smaller than the theoretical value of 7 $\mu_B$/Eu$^{2+}$. A fully polarized FM state maybe realized in higher magnetic fields. Furthermore, the positive MR between $H_c$ and $H_s$ at $T$ < 21 K might be the consequence of the competition between spin-related scattering in canted FM state (i.e., negative MR but less field dependent above $H_c$) and the orbit effect of charge carriers (i.e., positive MR but linearly increases with $H$). The negative MR between 21 K and 80 K provides strong indication of the magnetic fluctuation in this system.

### III. CONCLUSION

We have investigated systematically magnetotransport and magnetization studies on the high quality EuSn$_2$As$_2$ single crystals. We found two magnetization plateaus of ~ 5.6 $\mu_B$/Eu at 4.6 T and ~ 6.6 $\mu_B$/Eu at 10.0 T for $H//c$, respectively. The unusual magnetization plateaus are accompanied by the negative longitudinal MR and abnormal Hall resistance, indicating the complicated magnetic transitions from an AFM state to a canted FM state at $H_c$ and then to a polarized FM state at $H_s$

## ACKNOWLEDGMENTS

This work was supported by the National Key Research and Development Program of China, Grant No. 2016YFA0401003, 2017YFA0403502, the Natural Science Foundation of China (No. U19A2093, U2032214, U2032163, No. U1732274, No. 11904002), and Collaborative Innovation Program of Hefei Science Center, CAS (Grant No. 2019HSC-CIP 001), and Youth Innovation Promotion Association of CAS (Grant No.2017483), Natural Science Foundation of Anhui Province (No.1908085QA15).